\documentclass[aps,pre,twocolumn,showpacs,superscriptaddress,groupedaddress,amsmath,amssymb]{revtex4-1}
\usepackage{epsfig}
\usepackage{graphics}
\usepackage{lipsum}
\usepackage{sidecap}
\usepackage{graphicx} %Include figure files
\usepackage{dcolumn} %Align table columns on decimal point
\usepackage{bm}% bold math
\usepackage{sidecap}  
\begin{document}
\widetext

‮
\title{Mechanism for intensity induced chimera states in globally coupled oscillators}

\author{V.~K.~Chandrasekar$^{1}$}
\author{R.~Gopal$^{2,3}$}
\author{A.~Venkatesan$^{3}$}
\author{M.~Lakshmanan$^2$}
 
\affiliation{
$^{1}$Centre for Nonlinear Science\& Engineering, School of Electrical \& Electronics Engineering, SASTRA University, Thanjavur- 613 401, India.\\
$^{2}$Centre for Nonlinear Dynamics, School of Physics, Bharathidasan University, Tiruchirapalli-620024, India\\
$^{3}$Department of Physics, Nehru Memorial College, Puthanampatti,
Tiruchirapalli 621 007, India.
}

\date{\today}
            
\begin{abstract}
We identify the mechanism behind the existence of intensity induced chimera states in globally coupled oscillators. We find that the effect of intensity in the system is to cause multistability by increasing the number of fixed points. This in turn increases the number of multistable attractors and we find that their stability is determined by the strength of coupling . This causes the coexistence of different collective states in the system depending upon the initial state. We demonstrate that intensity induced chimera is generic to both periodic and chaotic systems. We have discussed possible applications of our results to real world systems like the brain and spin torque nano oscillators. 
\end{abstract}


\maketitle

\section{Introduction}
Dynamics of globally coupled oscillators continues to be a highly active topic of research for the past four decades or so because of its applicability to complex systems \cite{Kuramoto:84,Winfree:01,Pikovsky:01}. In particular, chimera is a dynamical state where synchronized and desynchronized oscillators coexist in networks of coupled identical oscillators \cite{Kuramoto:96}. Ranging from chemical reactions \cite{Tinsley:12}, laser arrays, and nano-oscillators to the cognitive behaviour of the brain \cite{Barbara:99}, chimera states play a crucial role in explaining various important dynamical behaviour involved in these real world systems \cite{Panaggio:14}. 
For instance, in the case of epileptic seizures, certain regions of the brain remain highly synchronized while the remaining regions are desynchronized \cite{Ayala:73}. In the case of Parkinson's disease, synchronized activity is absent in certain regions of the brain \cite{Levy:00} due to damaged or lost cells. In many mammals uni-hemispherical sleep is a phenomenon where only one hemisphere of the brain shows sleep activity and the sleeping side of the brain exhibits highly synchronized activity while the awaken side shows desynchronized activity \cite{Rattenborg:00}. Chimera states have also been identified or modeled in mechanical oscillator networks \cite{Martensa:14}, formation of ocular dominance stripes \cite{Swindale:80}, in social systems, during ventricular fibrillation \cite{Davidenko:92}, and so on \cite{Panaggio:14}. Very recently distinct measures to characterize these states have also been introduced \cite{Gopal:14}.

Earlier studies indicated that coupled oscillator systems should have weak, and nonlocal coupling in order to have chimera behaviour. Even though considerable promising results have been obtained both experimentally \cite{Tinsley:12, Hagerstrom:12, Martensa:14} and theoretically \cite{Abrams:06,Omelchenko:11}, the above two essential ingredients posed a restriction on the occurrence of chimera states.
Interestingly, recently chimera states have been found to exist in globally coupled systems as well. The occurrence of chimera states is found to be mediated by the amplitude in the coupling in complex Ginzburg-Landau systems \cite{Sethia:14}. The existence of chimera states due to delay induced multistability have also been recently reported \cite{Yeldesbay:14}.

In this paper, we explore and bring out explicitly the exact reason behind the occurrence of chimera states in globally coupled systems without the inclusion of weak, and nonlocal coupling. We find that the introduction of intensity dependent self interaction increases the number of fixed points which in turn increases the number of multistable attractors in the system whose stability is determined by the strength of the coupling between the individual units. On this reasoning we find that the system exhibits a multistable behaviour where it follows a path either to a synchronized state or to a chimera state depending upon the initial state. In order to investigate such intensity induced chimera states and to study the associated multistable behaviour, we consider two different coupled nonlinear dynamical systems, namely (1) coupled van der Pol oscillators and (2) coupled R\"ossler systems, as prototypical models. We investigate their collective behaviour in the periodic regime (of both the systems) as well as the chaotic regime (of R\"ossler system) of individual units.

The plan of the article as follows. In the following section we investigate the collective dynamics of a system of coupled van der Pol oscillators in the presence of an intensity dependent  frequency/interaction to show the existence of chimera states. Section III describes analytical confirmation for the existence of intensity induced chimeras along with a study of multistability in coupled van der Pol oscillators. In Sec. IV we further demonstrate the occurrence of intensity induced chimeras in globally coupled  R\"ossler oscillators and multistability nature when the individual units are oscillating either  periodically or chaotically. We point out applications of the proposed results and our conclusions in Sec.V. In Appendix A we present the salient features of characterization of chimera and other collective states in terms of the newly introduced measure~\cite{Gopal:14}, namely strength of incoherence ($S$), while in Appendix B we present some further details of the collective states and multistable behaviour in the chaotic regime of  R\"ossler oscillators.

%%%
\begin{figure*}
\centering
\includegraphics[width=0.75\linewidth]{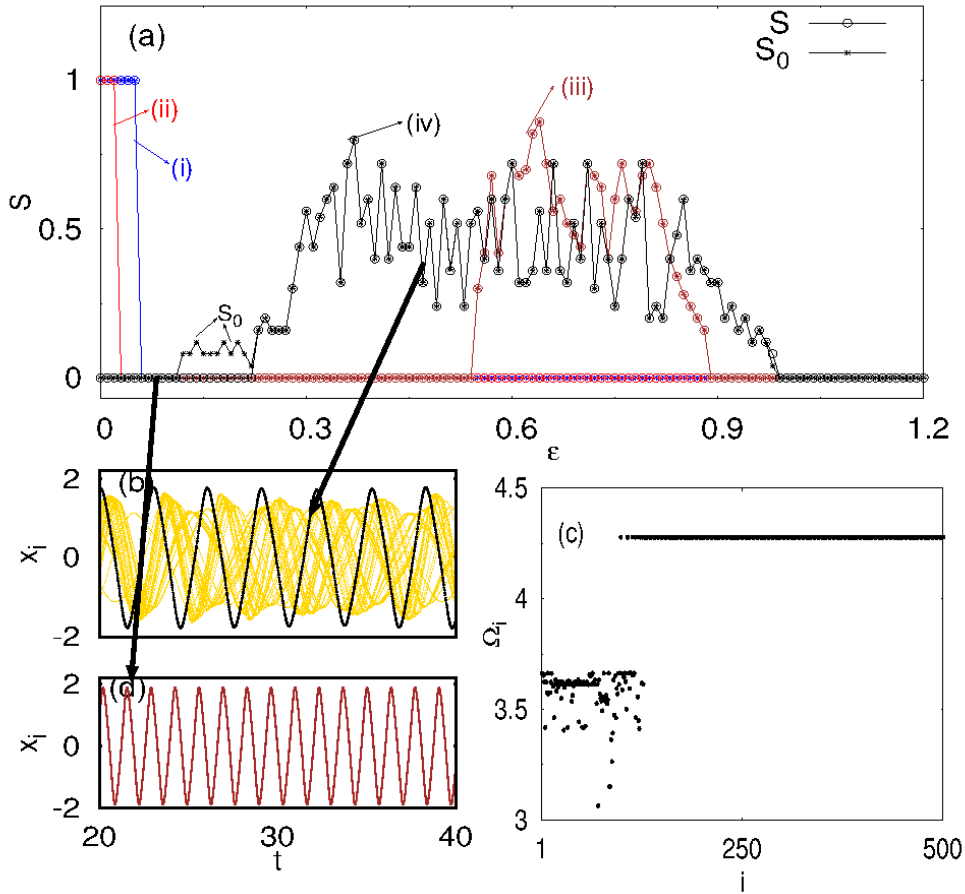}
\caption{(Color online) (a) The strength of incoherence $S$ (and also $S_{0}$ for cluster states) is plotted against coupling strength $\epsilon$ for different values of intensity parameters $\alpha_{1}, \alpha_{2}$ and coupling strength $\eta$ for system (1) for $N=500, b=1$ and $\omega_{0}=2.8$. The curves (i) - (iv) correspond to four different parametric choices: (i) $\alpha_{1}=\alpha_{2}=\eta=0$ (ii) $\alpha_{1}=2.15$, $\alpha_{2}=\eta=0$ (iii) $\alpha_{1}=4$, $\alpha_{2}=0$ and $\eta=0.1$ (iv) $\alpha_{1}=2.18$, $\alpha_{2}=2.15$ and  $\eta=0.1$. In the curve (iv) the range of $\epsilon$ where $S$ and $S_{0}$ are different correspond to cluster states.  (b) Time evolution of $x_{i}$ for a typical chimera state is plotted for the paramerers $\alpha_{1}=2.18$, $\alpha_{2}=2.15$, $\eta=0.10$, $\epsilon=0.42$ and $N=500$. Thick black line represents synchronized oscillations and the gold/grey lines represent desynchronized oscillations. (c) The averaged frequencies $\Omega_i=<\omega_{i}>$ of the chimera state shown in (b).  (d) Synchronized behaviour of the system for $\epsilon=0.10$ and other parameters as in (b).}
\end{figure*}
%%%

%%
\begin{figure}
\centering
\includegraphics[width=1.0\linewidth]{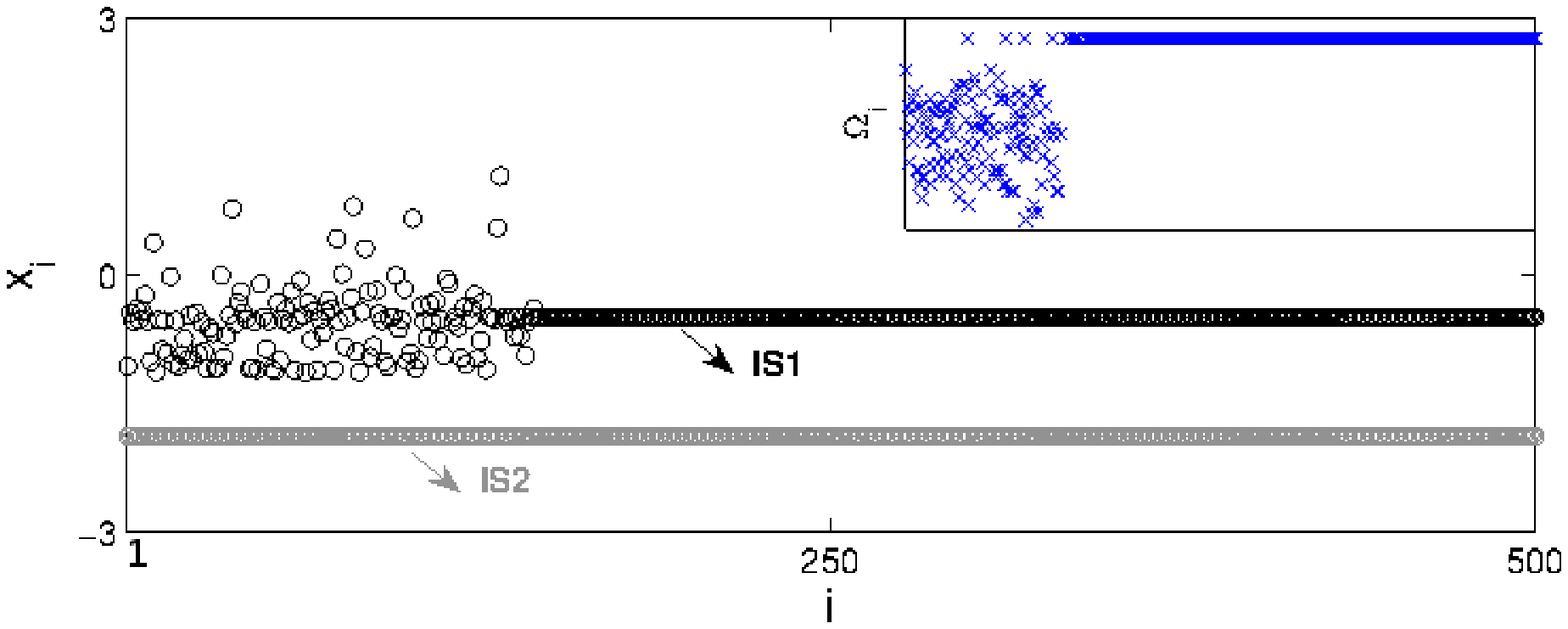}
\caption{\label{fig7} (Color online) Snapshots of $x_i$ are plotted for two different initial states IS1 (black) and IS2 (grey), where the system exhibits chimera and synchronized states in the coupled van der Pol system (1) with the parameters $\alpha_{1}=2.18$, $\alpha_{2}=2.15$, $b=1$, $\omega_{0}=2.8$, $\epsilon=0.62$ and $\eta=0.10$. Inset shows the corresponding time averaged frequencies $\Omega_i=<\omega_{i}>$ for IS1 (chimera state). Here, IS1 corresponds to a random distribution of $x_{i}'s$, while IS2 is close to a synchronized state.}
\end{figure}

\section{Intensity induced chimera in Coupled van der Pol Oscillators}

In order to exemplify our findings, we first consider a system of limit cycle exhibiting van der Pol oscillators coupled in a global fashion  with additional intensity dependent frequency terms,

\begin{eqnarray}
\ddot{x_{i}}&=&b(1-x_{i}^{2})\dot{x_{i}}-(\omega_0^2+\alpha_{1}x_{i}^{2}+\alpha_{2}x_{i}^{4})x_{i} \nonumber  \\
&&\quad+\epsilon(\dot{X}-\dot{x_{i}}) +\eta(X-x_{i}), \;i=1,..,N
\end{eqnarray}
where $\alpha_i$, $i=1,2$, are the intensity parameters, $\epsilon$ and $\eta$ denote the coupling strengths and $\{X,Y\}=\frac{1}{N}\sum_{i=1}^{N}\{x_{i},y_{i}\}$. It may be noted that Eq. (1) can also be designated as a system of coupled (external) force-free van der Pol-Duffing oscillators, when $\alpha_{2}=0$. We fix the parameter $b=1$ in Eq. (1), so that individual oscillators oscillate periodically in the absence of intensity dependent terms and couplings. Before proceeding further, we may note here that when the intensity dependent terms in  (1) are absent, that is $\alpha_{1}=\alpha_{2}=0$, one can identify chimera states~\cite{Omelchenko:11,Gopal:14} if one introduces nonlocal coupling of the form $\frac{\epsilon}{2P}\sum_{j=i-P}^{j=i+P}(x_{j}-x_{i})$ instead of the global coupling given in (1) between the oscillators. The present study is concerned with the role of intensity dependent terms $\alpha_{1}\neq0$, $\alpha
_{2}\neq0$ in realizing chimera states under global  coupling.   

To begin with, we demonstrate the need for the introduction of intensity dependent frequency/  interaction terms whose strengths are characterized by the coefficients $\alpha_{1}$ and $\alpha_{2}$ in the system (1) for  chimera states to exist. We consider the collective dynamics of $N=500$ oscillators specified by (1) for different choices of parameters. For our quantitative analysis, we use the recently introduced measure, namely strength of incoherence $S$~\cite{Gopal:14}, some details of which are indicated in Appendix A, to distinguish various collective states: (i) coherent state: $S=0$, (ii) chimera state: $0<S<1$ and (iii) ‫‫incoherent (desynchronized) state: $S=1$. We also distinguish cluster states from synchronized states by noting the fact that cluster states are essentially distinct groups of synchronized states characterized by finite number of discontinuities in the values of the variables $ x_{i}$. Introducing another measure $S_{0}$ which is the strength of incoherence before the removal of discontinuities in the difference variables $z_{i}=x_{i}-x_{i+1}$, while  the previous measure $S$ corresponds to the strength of incoherence after removal of discontinuity points by the method of removal of discontinuities~\cite{dis:92} (see ref.\cite{Gopal:14} for more details), we can easily see that for the cluster state $S_{0}$ takes a nonzero value ($0<S_{0}<1)$ while $S=0$. For other states, namely synchronized (coherent), desynchronized and chimera states, $S_{0}=S$, and so no disctinction is made in our further study between $S$ and $S_{0}$ for these states. Only for the cluster states both the quantities $S_{0}$ and $S$ are identified separately

Four different parametric choices are considered in our study and the resultant collective states characterized by the strength of incoherence $S$ (and $S_{0}$) are shown in Fig.  1(a). These four states correspond to the 
following. \\

(i) For the choice $\alpha_{1}=\alpha_{2}=\eta=0$, $\epsilon\neq0$, which corresponds to a system of coupled standard type van der Pol oscillators, we observe that the system transits from a desynchronized state, characterized by $S=1$, to a synchronous state ($S=0$) as the coupling strength $\epsilon$ is increased (Fig. 1(a)).\\

(ii) When the intensity coefficient $\alpha_{1}$ alone is increased to $\alpha_{1}=2.15$ while $\alpha_{2}=\eta=0$, the coupled oscillators again transit from a desynchronized state to a synchronous state as seen in Fig. 1(a).\\

(iii) Next, when we increase the value of the intensity coefficient $\alpha_{1}$ further (along with additional coupling $\eta=0.1$ for convenience), namely $\alpha_{1}=4$, $\alpha_{2}=0$, we observe that the system gets transited from a synchronized state ($S=0$) to a  chimera state ($0<S<1$), and then again to a synchronized state as the coupling strength $\epsilon$ is varied upwards.\\

(iv) Finally, on including  both the intensity dependent terms in (1) by choosing $\alpha_{1}=2.18$, $\alpha_{2}=2.15$, and keeping $\eta=0.1$, the system is found to again transit from a synchronized state to a chimera state via a cluster state $(0<S_{0}<1, S=0)$ and then again to a synchronized state (over a wider range in $\epsilon$) as $\epsilon$ is increased.

The above studies clearly demonstrate that chimera states result as a consequence of addition of intensity dependent self interaction. So we name this behaviour as {\it{intensity induced chimera states}}. In Fig. 1(b) we have presented the time evolution of a chimera state corresponding to a specific value of the coupling parameter $\epsilon=0.42$ for the choice (iv) above. We also confirm the chimera state nature in Fig. 1(b) by calculating the time averaged frequencies $\Omega_i=<\omega_{i}>$ in Fig. 1(c). Finally in Fig. 1(d) we present a synchronized state for the choice (iv) above for the specific choice $\epsilon=0.10$.

At this point, it also of considerable interest to note that the chimera states identified above exhibit a typical multistable behaviour depending on the initial conditions. In Fig. 2 we have presented the snapshots for the dynamical variables $x_{i}$ for two different initial states  IS1 (black) and IS2 (grey) in Fig. 2. In IS1 the initial states of the oscillators are randomly distributed in [-1,1], whereas in IS2 the initial states of the oscillators are chosen to be close to a synchronized state. For IS1 the system exhibits a chimera state, whereas for IS2 we see only a synchronized state. The nature of multistability is considered in more detail in the following section.                

\begin{figure}
\centering
\includegraphics[width=1.0\linewidth]{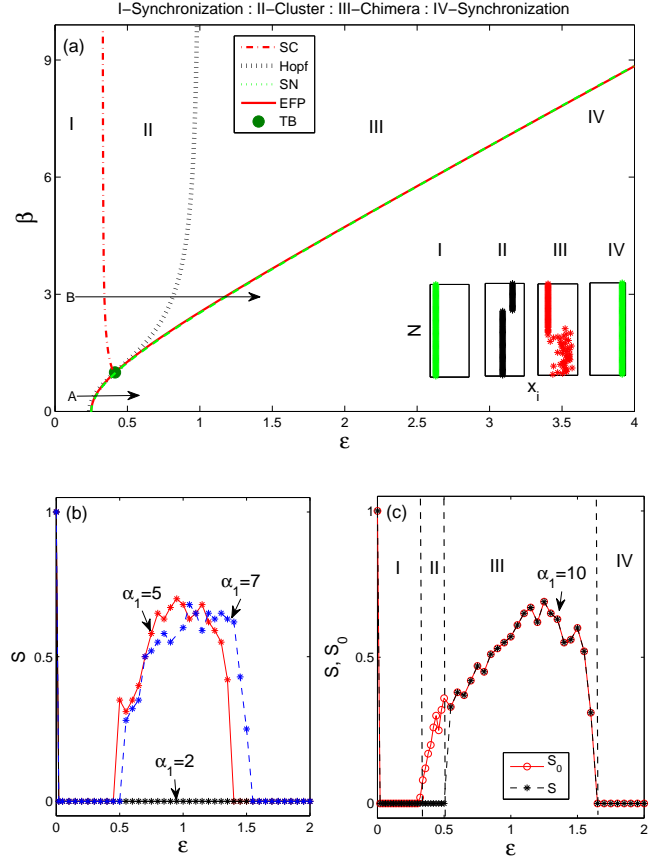}
\caption{(Color online) (a) Analytical bifurcation diagram in ($\epsilon$-$\beta$) space showing various dynamical regimes of system (2). (Here $\beta=3\alpha_{1}/\omega_{0}$). Dot dashed, dotted and dashed lines represent the saddle connection (SC), Hopf and saddle node (SN) bifurcation boundaries, respectively. Solid line and black circle represent the existence of fixed points (EFP) and Takens-Bogdanov point (TB), respectively. Regions I and IV represent synchronization regime and regions II and III represent the cluster and chimera regimes, respectively, corresponding to transition B. The insets show the snapshot of $x_i$ in the corresponding regions. Lines A and B represent two different synchronization routes taken by the system (explained in the text). (b), (c) The strengths of incoherence $S$ ($\star$) and $S_0$ $(\circ)$ are plotted against $\epsilon$ for various $\alpha_1$. The parameter values are $\omega_{0}=2.8$, $b=1$, $\zeta=0$ and $\gamma=0$.}
\end{figure}
%%%
\section{Multistability and transitions between different collective states: Analytical and Numerical Confirmations}

In order to understand the existence of above type of collective states and transitions between them as  the coupling strength $\epsilon$ and strength of intensity dependent coefficients ($\alpha_{1}$ and $\alpha_{2}$) are varied, we carry out an analytical investigation based on a multiple time scale approximation followed by a numerical study of the dynamical equation (1). We identify the role of multistability in the system as the main cause for the different transitions which occur in the collective behaviour as a function of the various parameters. Particularly we concentrate on the chimera state to start with. Using (1), in the chimera state, the synchronized $(x_s)$ and desynchronized $(x_d)$ oscillators can be represented by

\begin{eqnarray}
&&\ddot{x}_{(s,d)}=b(1-x_{{(s,d)}}^{2})\dot{x}_{(s,d)}-(\omega_0^2+\alpha_{1}x_{(s,d)}^{2} +\alpha_{2}x_{(s,d)}^{4})\nonumber  \\
&&\;\;\times x_{(s,d)}+\epsilon(\dot{X}-\dot{x}_{(s,d)})+\eta(X-x_{(s,d)})),
\end{eqnarray}
where $X=aX_s+cX_d$, $ s=1,2,\ldots l,  d=1,2,\ldots k $, \\
 $l+k=N, a=l/N$ and $c=k/N$.
The quantities $X_{s}$ and $X_{d}$ are given by 
\begin{subequations}
\begin{eqnarray}
\hspace{-0.2pt}&&X_{s}=\frac{1}{l}\sum_{s=1}^{l}x_{s}    \\
&&\hspace{-2.9cm}\mbox{and}~~~~~~~~~   \nonumber \\
\hspace{-0.2pt}&&X_{d}=\frac{1}{k}\sum_{d=1}^{k}x_{d}.    
\end{eqnarray}
\end{subequations}
In order to validate the numerical findings let us assume that the desynchronized oscillators are uncorrelated to each other and that they are only correlated to the synchronized group. In addition we assume that the influence of the desynchronized group on the synchronized group is negligible. This allows one to choose $c \approx0$, as chimeras emerge out from synchronized states only when the coupling strength $\epsilon$ is varied.  
 
\subsection{Multi-time scale approximation}
Let us apply a multi-time scale approximation (that represents the intensity induced slow variation in the amplitude and phase) 
\begin{equation}
x_{(s,d)}(t)=A_{(s,d)}(\tau) e^{i\omega_0 t_0}+A_{(s,d)}(\tau)^{*} e^{-i\omega_0 t_0},
\end{equation}
where $A_{(s,d)}(\tau)$ represents the amplitude of the synchronized ($s$) and desynchronized ($d$) oscillators and $t=t_0+\tau$; $t_o$ and $\tau$ represent the fast and slow time scales, respectively. Thus the dynamics of each synchronized and desynchronized oscillator is represented by

\begin{eqnarray}
\dot{A}_{s}=b A_{s}-(b+i(\beta+\gamma |A_{s}|^2))|A_{s}|^2 A_{s},
\end{eqnarray}

\begin{eqnarray}
\dot{A}_{d}=b A_{d}-(b+i(\beta+\gamma |A_{d}|^2))|A_{d}|^2 A_{d} \nonumber 
\\-(\epsilon+i\zeta)(A_{d}-A_{s}),
\end{eqnarray}
where $\beta=3\alpha_1/\omega_0,\;\gamma=5\alpha_2/\omega_0$ and $\zeta=-\eta/\omega_0$.

Then, the dynamics of the desynchronized oscillators can be rewritten as
\begin{eqnarray}
\frac{dA}{d\tau}&=&(b+i\Omega)A-(b+i(\beta+\gamma |A|^2))|A|^2A \nonumber  \\&&\qquad\qquad\qquad-(\epsilon+i\zeta )(A-1), 
\label{0}
\end{eqnarray}
where $A=A_d e^{i\Omega \tau}$, $A_s=e^{-i\Omega \tau}$ and $\Omega=\beta+\gamma$. Here $A_{s}$ is the stable solution of the synchronized oscillator.
Now the stability of the desynchronized state $A$ determines the stability of the chimera state. Eq. (\ref{0}) has five fixed points for $\gamma\neq0$ and three fixed points for $\gamma=0$ (implying $\alpha_{2}=0$). The specific fixed point $A=1$ is stable and represents the completely synchronized state of the system. The stability of the other fixed points (apart from the fixed point $A=1$) determine the existence of the chimera state.

In order to keep our analysis tractable, we will concentrate on the case $\gamma=0$ in the following.
One can note that from (\ref{0}) that for $\gamma=0$ the other two fixed points are given by
\begin{eqnarray}
&&\left|A\right|_{2,3}^{2}= \frac{1}{2 \left(b^2+\beta ^2\right)}(b^2+\beta ^2-2 b \epsilon -2 \beta \zeta \nonumber\\  &&\pm\sqrt{\left(b^2+\beta ^2-2 b \epsilon -2 \beta \zeta \right)^2-4 \left(b^2+\beta ^2\right) \left(\epsilon ^2+\zeta ^2\right)}).
\label{1}
\end{eqnarray}

We also note that these fixed points exist only when the expression inside the square root on the right hand side of (\ref{1}) is greater than or equal to zero. Consequently, the boundary in the $(\epsilon-\beta)$ plane for the existence of three fixed points is given by

\begin{eqnarray}
\epsilon_{f}=-(b^3+b \beta (\beta-2 \zeta)\pm(\left(b^2+\beta^2\right)^2 \left(b^2 \right. \nonumber \\   
\left. +\beta^2-4 \beta \zeta\right))^{\frac{1}{2}})/(2 \beta^2)   
\end{eqnarray}
In (9), we take into account only the positive solution as $\epsilon>0$ in our study.
Now, performing a linear stability analysis on equation (\ref{0}) we find that the trace of the Jacobian matrix  is zero ($\mbox{tr}(J)=0$) for any fixed point when

\begin{eqnarray}
\left|A\right|^{2}=\frac{b-\epsilon }{2 b}
\label{2}
\end{eqnarray}
and the determinant of the Jacobian matrix ($\mbox{det} (J)$) is zero for

\begin{eqnarray}
&&\left|A\right|^{2}=\frac{1}{2 \left(3 b^2+3 \beta ^2\right)}\bigg[4 b^2+4 \beta ^2-4 b \epsilon -4 \beta \zeta\nonumber\\ &&\pm\bigg(\left(-4 b^2-4 \beta ^2+4 b \epsilon +4 \beta \zeta \right)^2-4 \left(3 b^2+3 \beta ^2\right) \nonumber\\ &&\times\left(b^2+\beta ^2-2 b \epsilon +\epsilon ^2-2 \beta \zeta +\zeta ^2\right.)\bigg)^{\frac{1}{2}}\bigg].
\label{3}
\end{eqnarray}

Now from equations (\ref{1}) and (\ref{2}) we get a Hopf bifurcation curve $\epsilon_{H}$  given by 

\begin{eqnarray}
\epsilon_{H}=(-2 b^3+2 b \beta \zeta+(b^2 \left(5 b^4+\beta^3 (\beta-4 \zeta) \right.\nonumber  \\
\left. +2 b^2 \left(\beta^2-6 \beta \zeta-2 \zeta^2\right)\right))^{\frac{1}{2}})/(b^2+\beta^2) 
\end{eqnarray} 
and from equations (\ref{1}) and (\ref{3}) we get a saddle node bifurcation curve $\epsilon_{SN}=\epsilon_{f}$.

The stability nature of the fixed points is shown in Fig. 3(a) where we have plotted the bifurcation diagram in the $\epsilon$ vs $\beta$ space (Note that $\beta=3\alpha_{1}/\omega_{0}$). Dot dashed, dotted and dashed lines represent the saddle connection, Hopf and saddle node bifurcation boundaries, respectively.
The saddle connection curve is obtained numerically by solving Eq. (7). The filled circle in Fig. 3(a) represents the Takens-Bogdanov point which is obtained by solving two simultaneous equations, $\mbox{tr}(J)=0$
and $\mbox{det} (J)=0$ for the fixed points given by (8).

Further, from a detailed numerical analysis of (1), we confirm that in regions I and IV the system settle down in the synchronized state whereas in regions II and III the system exhibits clusters and chimera states, respectively. (See the insets of Fig. 3(a)).

The phenomenon of intensity induced chimera can now be better understood from the perspective of ($\epsilon$-$\beta$) space by looking at two specific transitions A and B in Fig. 3(a). Transition A denotes the case where the strength of  intensity  dependent term $\beta(=3\alpha_{1}/\omega_{0})$ is very low in the system and one can clearly see a synchronization-synchronization transition for increasing coupling strength (in fact one cannot name this a transition, but we want to emphasize the absence of other states). On the other hand, in the case of transition B for a higher strength of intensity ($\beta=5$)  the system takes a synchronization-cluster-chimera-synchronization transition route which is similar to the swing-by mechanism addressed by Daido and Nakanishi \cite{Daido:07}. However, in our case we find the existence of chimera state in addition to cluster states while the system swings by. The corresponding numerical behaviour of transition B is shown in the insets where we have plotted the snapshots of oscillator states in each region by solving Eq. (2).

The swing-by mechanism induced in system (2) is illustrated using the strength of incoherence before and after the removal of discontinuity \cite{Gopal:14}, $S_0$ and $S$, respectively in Figs. 3(b) and 3(c). The existence of synchronized, cluster, chimera and desynchronized states are represented by $S$ and $S_0$. As mentioned in Sec.II, ($S,S_0)=(1,1)$ represents a desynchronized state, while ($S,S_0)=(0,0)$ represents a synchronized state. Further, $S=0, 0<S_0<1$ and $0<S=S_0<1$ represent cluster and chimera states, respectively \cite{Gopal:14}. In Fig. 3(b) we have plotted S for different values of intensity $\alpha_1$. For $\alpha_1=5$ and $7$ we clearly see the existence of chimera states sandwiched between two synchronized states ($S=0$). However for a lower intensity, $\alpha_1=2$, the chimera state is absent and the system exhibits only a synchronized state as denoted by route A shown in Fig.  3(a). In Fig. 3(c) we have plotted both $S$ and $S_0$ for $\alpha_1=10$ and we can clearly see the existence of cluster states in region II where there is a mismatch between $S$ and $S_0$ followed by chimera states in region III. This is a realization of the route B indicated in Fig. 3(a). 

We also note here that considering equation (1) and carrying out a slow amplitude analysis by assuming 
$x_{i}(t)=A_{i}(\tau)e^{i \omega_0 t} +$ c.c, where $\tau$ represents the slow time scale, and working out a multi-time scale approximation we arrive at  the form $\dot{A_{i}}=(b+i\Omega)A_i-(b+i(\beta+\gamma |A_i|^2))|A_i|^2A_i-(\epsilon+i\zeta)(\bar{A}-{A_{i}}), \;i=1,..,N $, $\Omega=\beta+\gamma$, $\beta=3\alpha_1/\omega_0,\;\gamma=5\alpha_2/\omega_0$ and $\zeta=-\eta/\omega_0$, which is similar to the coupled Ginzburg-Landau oscillators equation (1) in ref. [18] when $\gamma=0$ and $b=1$. By following the above analysis for synchronized and desynchronized states, vide Eqs. (5) and (6), we have also confirmed that a similar scenario of chimera onset arises in the case of the coupled complex Ginzburg-Landau oscillators discussed in ref. [18].

%%%
\begin{figure}
\centering
\includegraphics[width=1.0\columnwidth=1.0]{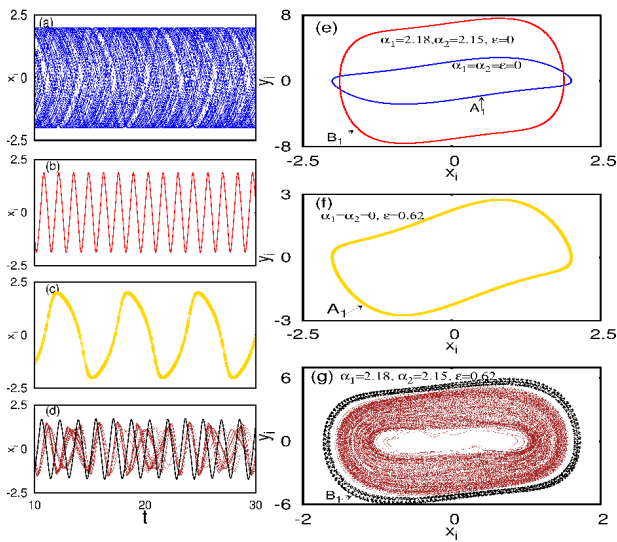}
\caption{Time series of system (2) for (a) $\alpha_{1}=\alpha_{2}=\epsilon=0$, (b) $\alpha_{1}=2.18, \alpha_{2}=2.15, \epsilon=0$, (c) $\alpha_{1}=\alpha_{2}=0, \epsilon=0.62$ and (d) $\alpha_{1}=2.18, \alpha_{2}=2.15, \epsilon=0.62$. (e) Phase plots of (a),(b). Figures (f) and (g) are the  phase plots corresponding to Figs. (c) and (d), respectively. Other parameters are same as in Fig. 1.}
\end{figure}

\subsection{Multistability : van der Pol Oscillators}

If we delve into the reasoning behind the occurrence of chimeras, one can deduce from the above analytical findings that the system acquires multistability upon introduction of intensity dependent interaction and sufficiently strong coupling. The effect of intensity dependent interaction in the system is to increase the number of fixed points (See. Eq. (6) and the discussion following it) whereas the stability of these fixed points is essentially determined by the coupling strength, see Fig. 3(a).

In our study we consider the synchronized oscillators which are uncoupled from the desynchronized ones so that the dynamics of the former is not influenced by the latter in Eq. (2), see also Eq. (5). We obtain the existence of multistability in the system and the reasoning behind the occurrence of chimera in Fig. 4. Initially in the absence of coupling $(\epsilon=0)$ and intensity terms $(\alpha_{1}=\alpha_{2}=0)$ we have plotted the time series of system (2) in Fig. 4(a) where all the oscillators in the system oscillate periodically with same amplitude and frequency but with different phases. Fig. 4(e) shows the phase plot for the system  corresponding to this desynchronized state as $A_{1}$.  Now, if we introduce intensity dependent terms $\alpha_{1}=2.18,\alpha_{2}=2.15$, the number of fixed points increases (in fact to five fixed points as pointed out above); however, only one stable synchronized attractor $B_1$ exists as shown in Fig. 4(e) red/grey circle. The corresponding time series of synchronized oscillations is shown in Fig.4 (b).

Now if we consider the coupling ($\epsilon \neq 0$) and the absence of intensity dependent terms ($\alpha_{1}=\alpha_{2}=0$) in the system, we find that again the oscillators are completely synchronized. This is also confirmed from the time series in Fig. 4 (c) and the corresponding  limit cycle (gold/grey)  in Fig. 4 (f) shows the existence of the synchronized attractor $A_1$ for this case. However in the presence of coupling and intensity dependent terms,  the coexistence of desynchronized and synchronized  attractors (Fig. 4(g)) demonstrates the existence of chimera state in the system which is evident from the corresponding time series shown in Fig. 4 (d). The bold black line represents the synchronized group in the attractor while the desynchronized oscillators  are shown in brown/grey lines. The corresponding attractor shown in the phase plot Fig. 4(g) clearly identifies the synchronized and desynchronized groups, and confirms the existence of a multistable attractor in the presence of chimera states.

\section{Intensity induced Chimeras in a systems of coupled R\"ossler oscillators}

In order to confirm the ubiquitous nature of the intensity induced chimeras we next consider a system of globally coupled R\"ossler oscillators:

\begin{eqnarray}
&&\dot{x}_{i}=-\omega_{0}(1-\alpha~r_{i}^{2})y_{i}-z_{i}+\epsilon(X-x_{i}) \nonumber \\
&&\dot{y}_{i}=\omega_{0}(1-\alpha~r_{i}^{2})x_{i}+ay_{i}+\epsilon(Y-y_{i})  \nonumber \\
&&\dot{z}_{i}=b+(x_{i}-c)z_{i}+\epsilon(Z-z_{i}), \;i=1,..,N
\label{4}
\end{eqnarray}
where $\{X,Y,Z\}=\frac{1}{N}\sum_{i=1}^{N}\{x_{i},y_{i},z_{i}\}$ and $-\alpha r_{i}^{2}=-\alpha (x_{i}^{2}+y_{i}^{2})$ represents the intensity dependent modification of the frequency. The parameter $\alpha$ represents the strength of intensity and $\epsilon$ is the coupling constant. In this case also we demonstrate the need for the introduction of intensity dependent frequency/interaction term in the system for the chimera state to exist.

{\it{(a)~Periodic regime:}}  For this purpose, we have plotted the strengths of incoherence $S$ (and also $S_{0}$ for cluster states) in Fig. 5(a) against the coupling strength $\epsilon$ for $\omega_{0}=1$, $a=0.2$, $b=1.7$ and $c=5.70$ (the individual units of the array are in the periodic regime for these parameters) for various values of the intensity parameter $\alpha$. For $\alpha=0$, we find that the system transits from a desynchronized state, characterized by $S=1$, to a synchronized state ($S=0$) upon increasing the coupling strength (represented by dotted line in Fig 5(a)). A chimera state ($0<S_{0}=S<1$) does not exist in the system in this case. Now, let us introduce intensity dependent interaction in the system by setting $\alpha=0.02$. The system takes a chimera route to synchronized state from the desynchronized state upon increasing the coupling strength as shown in Fig. 5(a) as dashed line. Similarly chimera states emerge for $\alpha=0.1$ (solid line). We also note that during the transition from chimera to synchronized states, cluster states ($<S_{0}<1, S=0$) arise in these cases as shown in Fig. 5(a).

In Figs. 5(b) and (c) we have shown the time evolution and the time averaged frequencies of the oscillators which are  plotted for $\alpha=0.02$ and $\epsilon=0.07$. We again clearly see the coexistence of synchronized and desynchronized oscillators. Fig. 5(d) demonstrates the absence of chimera states when $\alpha=0$, where only a synchronized state exists.  We again find that the present system (13) also exhibits intensity induced chimeras under global coupling.

{\it{(b)~Chaotic regime:}} A similar study confirming the existence of chimera states, depending on intensity dependent interaction, when the individual units are oscillating chaotically has been performed. Further details are given in Appendix B.

%%%%
\begin{figure}
\centering
\includegraphics[width=1.0\columnwidth=1.8]{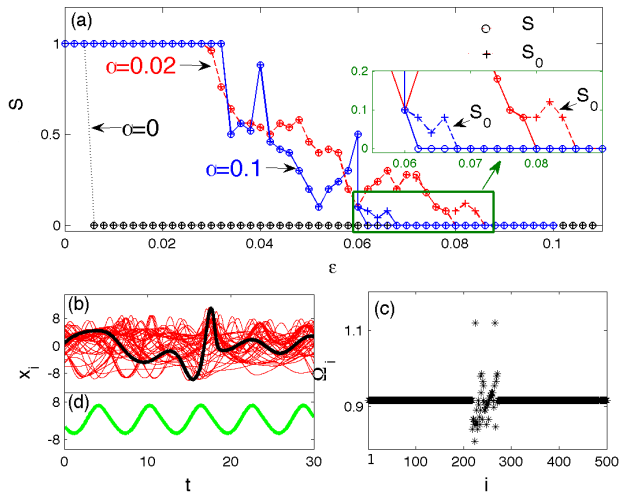}
\caption{\label{fig1} (Color online) (a) The strength of incoherence $S$ (and also $S_{0}$) is plotted against coupling strength $\epsilon$ for different values of intensity $\alpha$ for system (13). (b) Time evolution of $x_{i}$ for the chimera state is plotted for $\omega_{0}=1$, $\alpha=0.02$, $\epsilon=0.07$ and $N=500$. Thick black line represents synchronized oscillations and the red/grey lines represent desynchronized oscillations. (c) The averaged frequencies $\Omega_i=<\omega_{i}>$ of the chimera state  shown in (b).  (d) Synchronized behaviour of the system for $\alpha=0$ and other parameters as in Fig. (b).}
\end{figure}
%%%%%%

\subsection{Multistability: R\"ossler oscillators}
Let us validate again the claim for the existence of multistability nature by the introduction of intensity dependent interaction in the present system of coupled R\"ossler oscillators also, not only in the periodic regime but also in the chaotic regime of individual units as well. Considering (\ref{4}) the dynamics of the desynchronized oscillators $(x_{d},y_{d},z_{d})$ can be given by

\begin{eqnarray}
&&\dot{x}_{d}=-\omega_{0}(1-\alpha~r_{d}^{2})y_{d}-z_{d}+\epsilon(x_{s}-x_{d}) \nonumber \\
&&\dot{y}_{d}=\omega_{0}(1-\alpha~r_{d}^{2})x_{d}+ay_{d}+\epsilon(y_{s}-y_{d})  \nonumber \\
&&\dot{z}_{d}=b+(x_{d}-c)z_{d}+\epsilon(z_{s}-z_{d}), 
\label{5}
\end{eqnarray}

where $\{x_{s},y_{s},z_{s}\}$ are the state variables of the synchronized oscillators which are uncoupled from the desynchronized ones so that the dynamics of the former is not influenced by the latter. We now proceed to show the existence of multistability in the system (\ref{5}) and establish the reasoning behind the occurrence of chimera states, when the individual units are oscillating either periodically or chaotically as shown in Figs. 6 and 7, respectively.

{\it{(a)~Periodic regime:}}  First we consider the case where the individual oscillators of the array are in the periodic regime with the choice of parameters as $\omega_{0}=1$, $a=0.2$, $b=1.7$ and $c=5.70$ in (\ref{5}). In the absence of coupling $(\epsilon=0)$ and intensity independent interaction $(\alpha=0)$ we have plotted the time series of system (\ref{5}) in Fig. 6(a), where all the oscillators oscillate periodically with same amplitude and frequency but with different phases. Fig. 6(b) shows the phase plot for the above system, where the black circle $A_{1}$ corresponds to the desynchronized state shown in Fig. 6(a). We have taken $10^2$ different initial conditions and as we can see all of them are drawn towards the desynchronized periodic attractor $A_1$ (black circle). Now, if we introduce intensity dependent interaction in the system ($\alpha=0.01$) the number of fixed points increases ( as large as 14 numbers); however, only one stable desynchronized attractor $B_1$ exists as shown in Fig. 6(b), red/grey circle. The corresponding time series of the system is shown in Fig. 6(c).

Now if we consider the presence of coupling in the system (but in the absence of intensity dependent term), we find that the oscillators are completely synchronized as evident from the time series plotted in Fig. 6(d). The black circle in Fig. 6(e) shows the existence of one synchronized attractor $A_1$ for this case. However, in the presence of coupling if the intensity dependent term is also present ($\alpha\neq0$), in addition to the desynchronized attractor $B_1$ we get a synchronized attractor $B_2$. The attractor $B_1$ in the case of absence of coupling (shown in Fig. 6(b)) represents the oscillators that are desynchronized only in the phase but have the same amplitude and frequency. On the other hand, the attractor $B_1$ in the presence of coupling (shown in Fig. 6(e)) represents the oscillators that are not only desynchronized in phase but also have different amplitudes and frequencies. The coexistence of attractors $B_1$ and $B_2$ demonstrates the existence of chimera states in the system which is evident from the corresponding time series shown in Fig. 6(f). The bold green/black line represents the synchronized group in the attractor $B_2$ and the desynchronized oscillators in attractor $B_1$ are shown in thin red/grey lines.

{\it{(b)~Chaotic regime:}}  A similar scenario of existence of multistability can be realized in the chaotic regime of individual R\"ossler oscillator units of (\ref{4}) and (\ref{5}) for the  parameters chosen as $\omega_{0}=1$, $a=0.42$, $b=2$ and $c=4$, see also Appendix B.  In this case the individual oscillators oscillate chaotically, in the absence of coupling and intensity dependent interaction  ($\epsilon=\alpha=0$), and the corresponding time series and phase trajectories are plotted in Figs. 7(a) and 7(b), respectively. Now, if we introduce the intensity dependent interaction alone ($\alpha=0.02, \epsilon=0$), the system (\ref{5}) oscillates periodically with the same amplitude but different phases (Fig. 7(c)). In Fig. 7(b) the corresponding  desynchronized  attractor  is denoted as a closed curve $B_{1}$. On the other hand, if we introduce the coupling alone ($\alpha=0,  \epsilon=0.13$) without the intensity dependent term, we find that the chaotically evolving oscillators are completely synchronized as seen from the time series plotted in Fig.  7(d). The gold/grey attractor in Fig. 7(e) shows the corresponding synchronized chaotic attractor $A_{1}$. Finally, in the presence of both the coupling and intensity dependent interactions ($\alpha=0.02, \epsilon=0.13$),  the  chimera state emerges. This  is evident from the corresponding time series and phase plots which depict the coexistence of synchronized (blue/grey) and desynchronized (gold/grey) chaotic oscillations as shown in Figs. 7(f) and 7(e), respectively.

\begin{figure}
\centering
\includegraphics[width=1.0\linewidth]{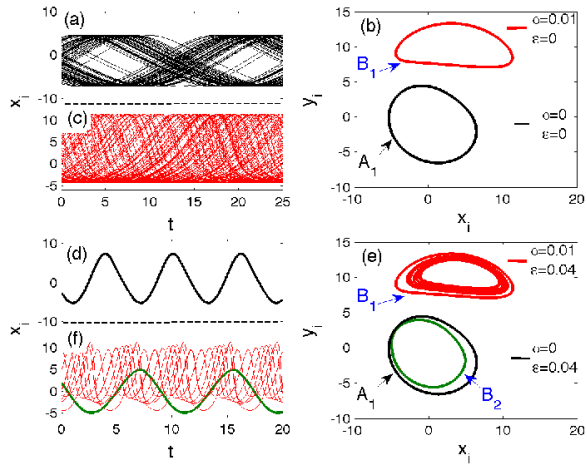}
\caption{(Color online) Time series of system (\ref{5}) for (a) $\alpha=0,\epsilon=0$ (c) $\alpha=0.01,\epsilon=0$, (d) $\alpha=0,\epsilon=0.04$ and (f) $\alpha=0.01,\epsilon=0.04$. (b) and (e) are the corresponding phase plots. Details are in text. Other parameter values are the same as in Fig. 5(b) (Here, individual units are oscillating periodically when $\epsilon=0$, $\alpha=0$).}
\end{figure}

\section{Conclusion}
From the above we conclude that even though the presence of intensity dependent terms in globally coupled system increases the number of fixed points, their stability is determined by the strength of coupling. Since there are more number of stable attractors in the system when coupling is present (in both the van der Pol and R\"ossler cases), depending upon the initial state the system will settle down to a chimera state or a different collective state as demonstrated in Figs. 2 and 9 (Appendix B) where for two different initial states we get chimera and synchronized states. If we choose the initial state of the system to be close to attractor $B_2$ (Fig. 6(e)) we get a synchronized state in the system for increasing coupling strength. On the other hand, if we choose the initial state so that the oscillators are spread across close to different attractors (like Fig. 6(e)) we can either get a chimera state or a multiclustered state. 

The above dynamical aspects resemble task specific synchronization/desynchronization in the brain \cite{Jane:09,Pfurtschelle:99}, where depending upon the state of the neuronal oscillations at the time of incoming tasks, the brain either chooses to accept a new task or ignores it and continues with the current task. Thus chimera states play a crucial role in the functioning of a normal brain where it is capable of taking up and processing multiple tasks at once (corresponding to multistability). On the other hand, in the case of pathological brain, due to the presence of damaged cells or due to the loss of normal brain cells \cite{Diesmann:99}, the coupling in the system is either too low or too high (as experienced by the individual neurons) so that the system swings to a mass pathological synchronization state corresponding to regions I or IV in Fig. 3.

In the case of spin torque nano oscillators the problem of coupling a large number of oscillators in order to achieve coherent microwave power still remains open \cite{Mohanty:05}. This is because introducing the coupling leads to an increase in the number of stable attractors in the system \cite{Zhou:12}, and this causes the emergence of chimera like states (synchronized clusters and desynchronized oscillators coexist) \cite{Subash:14}. 

Our results therefore help in identifying the underlying cause for the crucial dynamical phenomena that occur in real world systems. The results also assist in gaining a better understanding of those dynamical phenomena thereby possibly helping researchers to control the occurrence of them depending upon whether they are desirable (task specific synchronization in the brain, and synchronization of spin torque nano oscillators) or undesirable (pathological mass synchronization).

\section{Acknowledgments}
The work of R.G. and M.L has been supported by the Department of Science 
and Technology (DST), Government of India sponsored IRHPA research project. M.L. has also been supported by a DAE Raja Ramanna fellowship and a DST Ramanna program.

\begin{figure}
\centering
\includegraphics[width=1.0\linewidth]{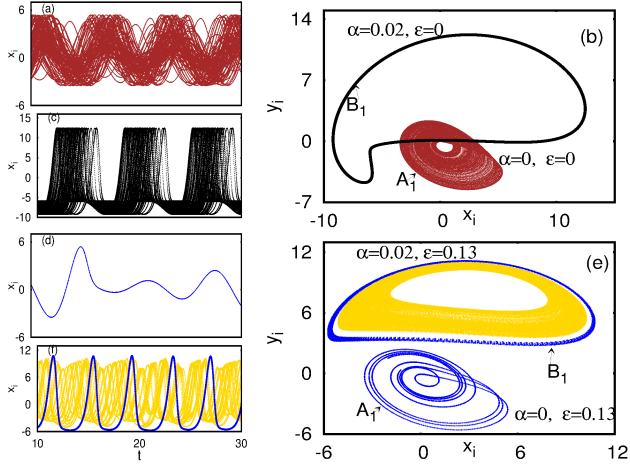}
\caption{(Color online) Time series of system (\ref{5}) for (a) $\alpha=0,\epsilon=0$ (c) $\alpha=0.02,\epsilon=0$, (d) $\alpha=0,\epsilon=0.13$ and (f) $\alpha=0.02,\epsilon=0.13$. (b) and (e) are the corresponding phase plots.  Other parameter values are $\omega_{0}=1$, $a=0.42$, $b=2$ and $c=4$ (the individual nodes are oscillating in the chaotic states for these parameters).}
\end{figure}

\begin{appendix}
\section*{Appendix A: Quantitative measure: Strength of Incoherence}
Recently we have introduced a specific quantitative measure, namely the strength of incoherence ($S$)~\cite{Gopal:14} to represent various dynamical states. To understand the quantitative measure $S$, we consider the  difference dynamical variables of the system $z_{i}=x_{i}-x_{i+1}$, $i=1,2,...,N$. Now, the occurrence of different synchronized states in the coupled dynamical system can be quantified by using the standard deviation  given by  
\begin{equation}
\sigma_{l}=\Big<\noindent \sqrt{\frac{1}{N}\sum_{i=1}^{N}[z_{i}-<z>]^2} \hspace{0.1cm} \Big>_{t},
\end{equation}
where  $i=1,2...N$, and  $<z>=\frac{1}{N}\sum_{i=1}^{N}z_{i}(t)$ and $\langle ...\rangle_{t}$ denotes average over time. Consequently $\sigma_{l}$'s take a value zero for synchronized states and nonzero values for both desynchronized  and chimera states.  To overcome the latter difficulty we divide the oscillators into $M$ (even) bins of equal length $n=N/M$. Then, we introduce the local standard deviation $\sigma_{l}(m)$ which can be defined as
\begin{equation}
\sigma_{l}(m)=\Big<\noindent \sqrt{\frac{1}{n}\sum_{j=n(m-1)+1}^{mn}[z_{j}-<z>]^2} \hspace{0.1cm}\Big>_{t}, \hspace{0.1cm}m=1,2,...M.
\label{6}
\end{equation}

The above quantity $\sigma_{l}(m)$  is calculated for every successive $n$ number of oscillators. Using (\ref{6}) we can introduce the measure named strength of incoherence $S$ as \\

\begin{equation} 
S=1-\frac{\sum_{m=1}^{M}s_{m}}{M}, \hspace{0.1cm}  s_{m}=\Theta(\delta-\sigma_{l}(m)),
\end{equation}
where $\Theta(.)$ is the Heaviside step function, and $\delta$ is a predefined threshold that is reasonably small. Here, we take $\delta$ as a certain percentage value of difference between $x_{i,max}$ and $x_{i,min}$.  Thus when $\sigma_{l}(m)$ is less than $\delta$, the value of $s_{m}=1$, otherwise it is '0'. Consequently, $S$  takes the values $S=1$ or $S=0$ or  $0 < S < 1$ for desynchronized, synchronized and chimera states, respectively. The cluster states can also be distinguished, as discussed  in Sec. II, by introducing another strength of incohernce $S_{0}$. For details see ref~\cite{Gopal:14}.

\begin{figure}
\centering
\includegraphics[width=1.1\linewidth]{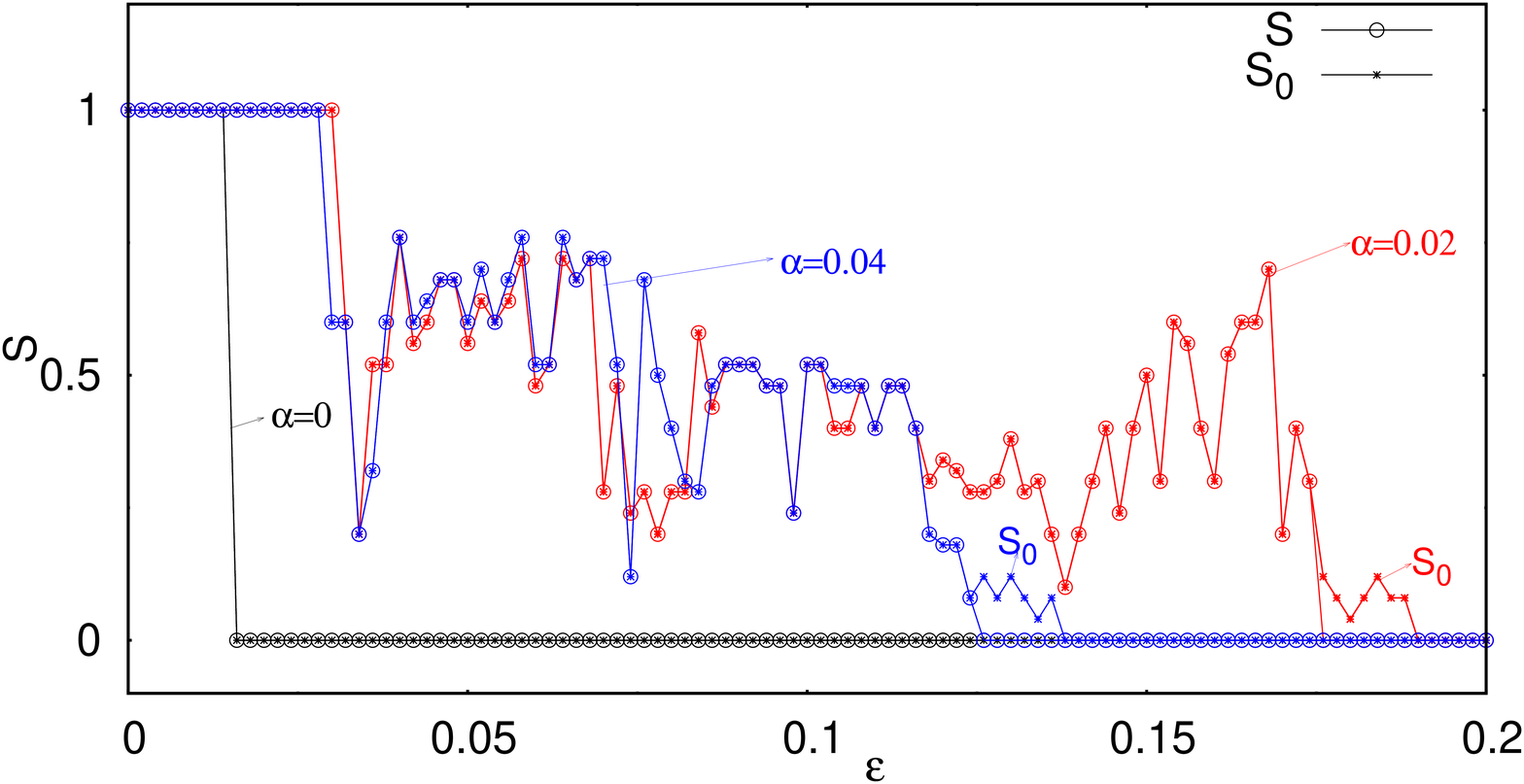}
\caption{\label{fig1} (Color online) The strength of incoherence $S$ (and also $S_{0}$) is plotted against coupling strength $\epsilon$ for different values of intensity $\alpha$ for system (\ref{4}) for $N=500$ when the individual units are oscillating in the chaotic regime. The parameter values are $\omega_{0}=1$, $a=0.42, b=2$ and $c=4$. The values of the other parameter $\alpha$ are indicated in the figure.}
\end{figure}

\section*{Appendix B: Chimeras and Multistable Behaviour in the Chaotic Regime of Coupled  R\"ossler Oscillators}

In this Appendix, we present the various collective states of (\ref{4}) when the individual  R\"ossler oscillators, in the absence of intensity dependent interaction ($\alpha=0$) and coupling ($\epsilon=0$), evolve chaotically. We have plotted in Fig. 8 the strength of incoherence  $S$ (and also $S_{0}$ for cluster states) as a function of coupling strength $\epsilon$, for various values of the intensity parameter $\alpha$. The other parameters are chosen as $\omega_{0}=1$, $a=0.42,  b=2$ and $c=4$ as in Fig. 7. For $\alpha=0$, the system transits from a desynchronized state ($S=1$)  to a synchronized state ($S=0$). Next, when the intensity dependent interaction $\alpha$ is included as $\alpha=0.02$, the system transits from a desynchronized state to a chimera state ($0<S=S_{0}<1$) and then transits to a synchronized state as $\epsilon$ is increases. We may also note that during the transition from chimera state to synchronized state, cluster states also occur and they are characterized by $S=0$, $0<S_{0}<1$. Similar behaviour occurs for $\alpha=0.04$ also.

Further, we have also confirmed that as in the case of van der Pol oscillators (Fig. 2) the existence of chimera states in the coupled R\"ossler systems (11) is essentially due to the presence of intensity dependent term ($\alpha\neq0$) and it exhibits multistable behaviour depending on the initial state even when the individual units are evolving in the chaotic regime as shown in Fig. 9. In Fig. 9, we have plotted the snapshots of dynamical variables $x_{i}$ fo two different initial states IS1 (oscillators are randomly distributed in the range [-1,1]) and IS2 ( all the units are close to a synchronized state). For IS1 the system exhibits a chimera state, while the system evolves into a synchronized state for IS2. The time averaged frequency for IS1 is also plotted in the inset of Fig. 9 which also confirms the above facts. From Fig. 9, it is clear that the system typically exhibits a multistable behaviour.

\begin{figure}
\centering
\includegraphics[width=1.00\columnwidth=0.8]{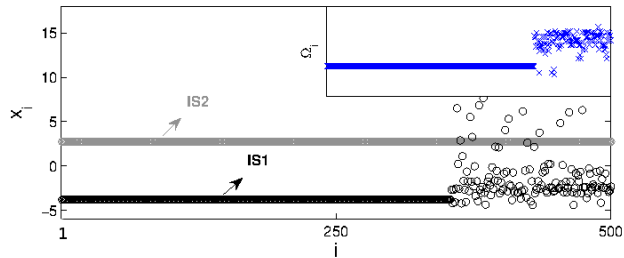}
\caption{\label{fig8} (Color online) Snapshots of $x_i$ are plotted for two different initial states IS1 (black) and IS2 (grey), where the system exhibits chimera and synchronized states in  the coupled R\"ossler system (\ref{4}) with the parameters chosen as $\omega_{0}=1$, $\alpha=0.02$, $\epsilon=0.13$, $a=0.42$, $b=2$ and $c=4$ (chaotic regime of individual units). Inset shows the corresponding time averaged frequency $\Omega_i=<\omega_{i}>$ for IS1 (chimera state).}
\end{figure}

\end{appendix}

\end{document}